\documentclass[aps,prl,preprint,groupedaddress]{revtex4} 
\usepackage{amsfonts}
\usepackage{amsmath}
\usepackage{graphicx,epsfig}
\usepackage{color}
\usepackage[
            colorlinks=false,bookmarks]{hyperref}
\usepackage{mathptm}
\usepackage{amssymb}
\usepackage{graphicx}

\newcommand{\di}{\partial}

\begin{document}

\title{A quantum spin transducer based on nano electro-mechancial resonator arrays}
\author{P.~Rabl$^{1}$}
\author{S.~J.~Kolkowitz$^{2}$}
\author{F.~H.~Koppens$^{2}$}
\author{J.~G.~E.~Harris$^{3}$}
\author{P.~Zoller$^{4}$}
\author{M.~D.~Lukin$^{1,2}$}
\affiliation{$^1$ITAMP, Harvard-Smithsonian Center for Astrophysics, Cambridge, MA 02138, USA}
\affiliation{$^2$Department of Physics, Harvard University, Cambridge, MA 02138, USA}
\affiliation{$^3$Department of Physics and Applied Physics, Yale University, New Haven, CT, USA}
\affiliation{$^{4}$Institute for Theoretical Physics, University of Innsbruck, and
Institute for Quantum Optics and Quantum Information of the Austrian Academy
of Science, 6020 Innsbruck, Austria}
\date{\today }

\begin{abstract}
\textbf{
Implementation of quantum information processing faces the contradicting requirements of combining excellent isolation to avoid decoherence with the ability to control coherent interactions in a many-body quantum system. For example, spin degrees of freedom of electrons and nuclei provide a good quantum memory due to their weak magnetic interactions with the environment. However, for the same reason it is difficult to achieve controlled entanglement of spins over distances larger than tens of nanometers. Here we propose a universal  realization of a quantum data bus for electronic spin qubits where spins are coupled to the motion of magnetized mechanical resonators via magnetic field gradients. Provided that the mechanical system is charged,  the magnetic moments associated with spin qubits can be effectively amplified to enable a coherent spin-spin coupling over long distances via Coulomb forces. Our approach is applicable to a wide class of electronic spin qubits which can be localized near the magnetized tips and can be used for the implementation of hybrid quantum computing architectures. 
}
\end{abstract}

\maketitle
Motivated by the challenge of implementing quantum information processing in real physical systems a wide variety of approaches are currently being explored in which the quantum bits stored in long lived states, such as those associated with spins, are mapped into other degrees of freedom to enable strong long-range coupling. This is frequently done by employing a quantum data bus that is specific to each particular qubit realization, for example mapping spin qubits to harmonic motion of trapped ions~\cite{CiracZoller,LeibfriedNAT2005,HaenselNAT2005} or photonic states associated with optically active qubits like atoms~\cite{MaunzNAT2007} or nitrogen-vacany (NV) centers in diamond~\cite{NVReview,JelezkoPRL2004,HansonPRL2006,ChildressScience2006,DuttScience2007}. However, such channels are absent for many spin qubits including prominent examples such as phosphor donors in silicon~\cite{KaneNature1998,MortonNAT2008,TyryshkinPRB2003,StegnerNatPhys2006} or N$@$C$_{60}$~\cite{HarneitPRA2002, BenjaminJPhys2006,MortonNatPhys2006}, as well as new generations of carbon~\cite{ChurchillPreprint,TrauzettelNatPhys2007}, and silicon~\cite{VrijenPRA2000,Zwanenburg} based quantum dots.

\begin{figure}
\begin{centering}
\includegraphics[width=0.65\textwidth]{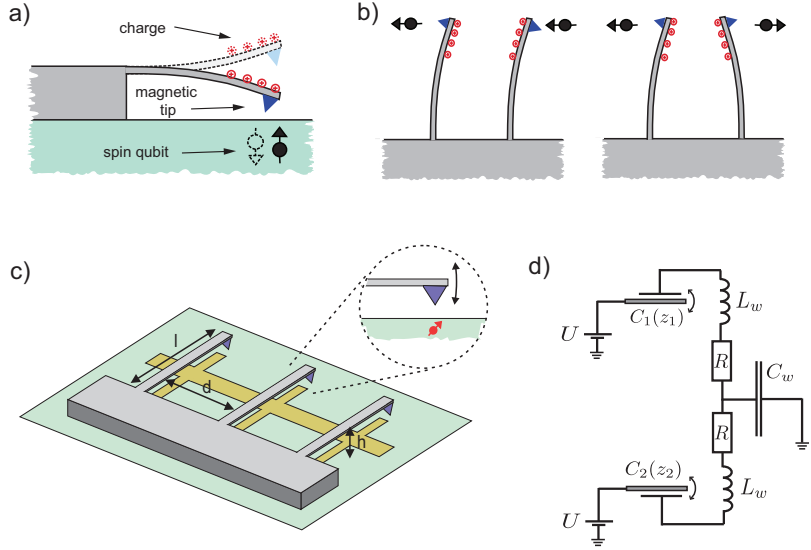}
\caption{{\bf Electro-mechanical quantum transducer.} {\bf a}, Magnetic interactions between a mechanical resonator and an electronic spin qubit lead to a spin dependent displacement of the resonator tip and convert the magnetic dipole of the spin into an electric dipole proportional to the charge on the resonator.  {\bf b}, For two resonators the difference in the electrostatic energies associated with different spin configurations then result in effective spin-spin interaction. {\bf c}, Schematic view of a spin register based on an electro-mechanical quantum bus. Here resonators are coupled indirectly via capacitive interactions with nearby wires to enable coupling of spins separated by $d\sim100\,\mu$m and to design spin-spin interactions by different circuit layouts.  {\bf d}, Circuit model which describes the coupling between two resonators. } \label{fig:Figure1}
\end{centering}
\end{figure}

Dramatic advances have recently been made in the fabrication and manipulation of micro and nano electro-mechanical systems (NEMS). Examples range from applications of NEMS as nanoscale magnetometers with single spin resolution~\cite{RugarSingleSpin} to cooling of individual mechanical modes close to the quantum ground state~\cite{NaikNAT2006,GiganNAT2006,SchliesserNATPHY2008,ThompsonNAT2008,TeufelPRL2008}.  In the following we show that such NEMS can be used to create a universal quantum transducer for spin-spin interactions. In our approach, illustrated in Fig.~\ref{fig:Figure1} a) and b), spin qubits are coupled to the motion of a magnetized NEMS via magnetic field gradients~\cite{PoggioNatNano,NVCantilever}. By application of an appropriate gate voltage the mechanical system is charged and the magnetic moments associated with spin qubits can be effectively amplified to enable coherent electric interactions over distances exceeding $100$ micrometers. The key advantages of our approach are that multiple spin setups can be designed and controlled using different electric circuit layouts in a scalable architecture, the ability to couple dissimilar spins to each other, and the potential to enable realization of hybrid systems in which spins are coupled to solid-state charge qubits~\cite{ArmourPRL2002} or isolated trapped atoms~\cite{TreutleinPRL2007} or ions~\cite{TianPRL,HensingerPRA2005}.

\section{Spin register with an electro-mechanical quantum bus}
In what follows we discuss the implementation and operation of a spin quantum register in which effective long range spin-spin interactions are mediated by an electro-mechanical quantum bus as shown in Figure 1 c). The system consists of an array of $N$ nano-mechanical resonators each coupled magnetically to an electronic spin qubit associated with an impurity located in the substrate below. The motion of each resonator tip $i=1,\dots,N$ along the z-axis is quantized and described by the Hamiltonian $H_r^i=\omega_r a^\dag_i a_i$ where  $a_i$ ($a_i^\dag$) are annihilation (creation) operators for the fundamental vibrational mode of frequency $\omega_r$.  We model each impurity as a $S=1/2$ electronic spin with Hamiltonian $H^i_{s}(t)= (\delta_i \sigma_z^i +\Omega_i(t) \sigma_x^i)/2$, where $\sigma_{x,y,z}^i$ are Pauli operators and $\Omega_i$ and $\delta_i$ are the Rabi frequency and detuning of local microwave fields. Strong magnetic field gradients $G_m$ produced by the magnetic tip result in a spin-resonator interaction $H_{sr}^i=  \frac{\lambda}{2} (a^\dag_i +a_i) \sigma_z^i$. Here $\lambda=g_s \mu_B G_m a_0/\hbar$, where $g_s\simeq 2$ and $\mu_B$ is the Bohr magnetron, is the Zeeman shift associated with the zero point motion $a_0=\sqrt{\hbar/2m\omega_r}$ of a resonator with vibrating mass $m$.  For Si nano-mechanical resonators with typical dimensions $(l,w,t)\approx (10,0.1,0.1)\,\mu m$ and frequencies $\omega_r/2\pi \sim 1$ MHz  we obtain $a_0\approx 3.5 \times 10^{-13}$ m and for a spin-tip separation $\lesssim 50$ nm the resulting coupling can approach $\lambda/(2\pi)\approx100$ kHz~\cite{NVCantilever}. Note that intrinsic spin coherence times $T_2$ in the range of 1-10 ms have been observed with  NV centers~\cite{ChildressScience2006,WrachtrupNatMat2009} or phosphor donors~\cite{TyryshkinPRB2003}. 

To establish long-range interactions between different sites the resonators are charged and interact capacitively with nearby wires interconnecting them. Variations of the resonator-wire capacitance $C_i(z_i)$ with the position of the tip $z_i=a_0(a_i+a_i^\dag)$  then introduce effective interactions between the resonators. For the length scales of interest electric resonance frequencies are large compared to mechanical frequencies and the phonon-phonon coupling $\hbar g_{ij} =  a_0^2 \frac{\di^2 W_{el}}{\di z_i \di z_j}|_{\{z_i\}=0}$ follows directly from the electrostatic energy $W_{el}$ of the underlying circuit. The resulting Hamiltonian for the coupled resonator array is
\begin{equation}
H_{\rm ph}\simeq \sum_i \omega_r a_i^\dag a_i + \frac{1}{2}\sum_{i, j} g_{ij}  (a_i+a_i^\dag) (a_j+a_j^\dag)=\sum_n \omega_n a_n^\dag a_n\,,
\end{equation}
where $\omega_n$ and $a_n=\sum_i c_{n,i} a_i$ denote frequencies and mode operators of collective phonon modes and we have absorbed a renormalization of the bare oscillation frequency into the definition of $\omega_r$. To  estimate the typical coupling strength we consider two sites separated by a distance $d$ and connected by a wire of self-capacitance $C_w\approx \epsilon_0 d$, as shown in Fig.~\ref{fig:Figure1} d). Then $W_{el}= -\frac{U^2}{2} \frac{ C_\Sigma C_w }{C_\Sigma + C_w}$, where $C_\Sigma = C_1(z_1)+C_2(z_2)$ and $U$ is the applied voltage.  Assuming $C_i(z_i)\simeq C(1-z_i/h)$, where $h$ is the mean electrode separation,
\begin{equation}
\hbar g =       \frac{ C^2 C_w^2 U^2}{(2C+C_w)^3} \frac{a^2_0}{h^2}\,.
\end{equation}
For resonator dimensions given above and choosing $h\approx w$,  i.e. $C\approx \epsilon_0 l$, we obtain $g/ (2\pi) \approx  95\, {\rm kHz}\times U[V]$ and for a doubly clamped beam voltages up to $U\approx10$ V~\cite{NaikNAT2006} lead to phonon-phonon interactions as large as 1 MHz. For a finite wire resistance $R$  dissipation of currents introduces an additional damping mechanism for the resonator motion which we describe by an effective Q-value, $ Q_{\rm el}  \approx  \omega_r m h^2/ U^2 C^2 R$. For a $\sim 200$ nm thick gold wire at a temperature $T=1$ K we obtain $R\sim 0.5$ Ohm. For parameters used above and voltages up to $U=10$ V we find that $ Q_{el}\gtrsim  10^7$ is above typical intrinsic mechanical Q-values and ohmic losses therefore do not impose a severe limitation on the coupling strength.

In summary we obtain the full spin register Hamiltonian $H=\sum_i H_{s}^i + \sum_i H_{sr}^i+ H_{ph}$ which by setting $\lambda_{n,i}=\lambda c_{n,i}$ is 
\begin{equation}\label{eq:Model}
H= H_{s}(t) + \sum_n \omega_n a_n
^\dag a_n + \frac{1}{2} \sum_{i,n} \lambda_{i,n} (a^\dag_n +a_n) \sigma_z^i.
\end{equation}
The characteristics of the electro-mechanical quantum bus appear in Eq.~\eqref{eq:Model} in the form of the phonon spectrum $\omega_n$ and the mode coefficients $c_{n,i}$, which in turn are determined by the electric circuit layout. As described in the following, this property offers a simple way to design and control different types of spin-spin interactions.

\section{Spin-spin interactions}
The model of spins coupled to a set of phonon modes as described by Eq.~\eqref{eq:Model} is familiar from quantum computing proposals with trapped ions~\cite{CiracZoller,SorensenPRL1999,MolmerPRL1999,Wunderlich2002,GarciaRipollPRL2003, GarciaRipollPRA2004} and in principle similar gate schemes as developed in this field can be applied for spin entangling operations here. To be compatible with the less favorable decoherence processes of the present physical implementation we focus on gate operations based on off-resonant spin phonon interactions~\cite{SorensenPRL1999,MolmerPRL1999,Wunderlich2002,GarciaRipollPRL2003, GarciaRipollPRA2004}. Such schemes avoid ground state cooling requirements, are consistent with spin echo techniques and allow a virtual elimination of motional dephasing processes in the limit of long lived spin qubits. Alternatively, gate operations based on a resonant exchange of phonons~\cite{CiracZoller} could be implemented using techniques described in Ref.~\cite{NVCantilever}.

\begin{figure}
\begin{centering}
\includegraphics[width=0.75\textwidth]{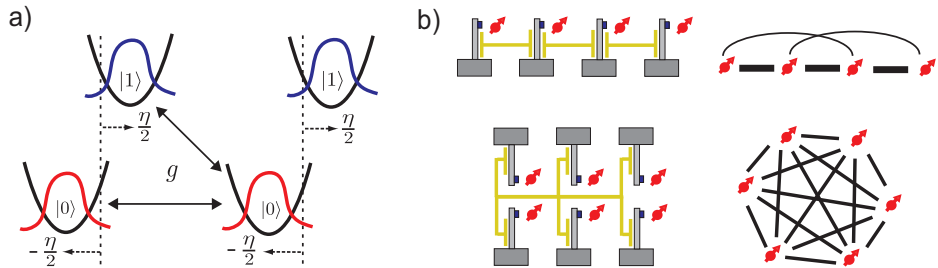}
\caption{{\bf Spin-spin interactions.} {\bf a}, In the `displaced oscillator basis' different spin states $|0\rangle$, $|1\rangle$ are associated with different resonator equilibrium positions displaced in phase space by an amplitude $\pm\eta/2$ where $\eta=\lambda/\omega_r$ is the magnetic coupling parameter.  Consequently, electric interactions between the resonators, $g\neq 0$, translate into Ising interactions $\sim\sigma_z^1\sigma_z^2$ with a strength proportional to $\eta^2$. {\bf b}, Two basic examples for different circuit layouts (left) and the corresponding effective Ising interactions (right), where the coupling strength $|M_{ij}|$ is indicated by the thickness of the lines.  } \label{fig:Figure2}
\end{centering}
\end{figure}

For the discussion of effective spin-spin interactions  it is convenient to change to a displaced oscillator basis which is related to the uncoupled basis by a polaron transformation  $U=e^{-i S}$,  where $S= \frac{1}{2} \sum_{i}  P_i\sigma_z^{i}$ and $P_i= i \sum_n  \lambda_{n,i}/\omega_n (a^\dag_n-a_n)$  are collective momentum operators.
In this new representation and for $\delta_i=0$ the resulting spin register Hamiltonian, $ H \rightarrow  UHU^\dag$,  is~\cite{Wunderlich2002}
\begin{equation}\label{eq:Hpolaron}
H= \sum_i \frac{\Omega_i(t)}{2}\left( \sigma_+^i e^{-i P_i} +  \sigma_-^i e^{i P_i}\right)   -\sum_{i\neq j} M_{ij} \sigma_z^{i}\sigma_z^{j} + H_{\rm ph} \,.
\end{equation}
Let us for the moment assume $\Omega_i(t)=0$ where Eq.~\eqref{eq:Hpolaron} reduces to an ``always-on" Ising interaction with coupling strength $M_{ij}=\sum_n \lambda_{n,i}\lambda_{n,j}/(4\omega_n)$,  which is mediated by but independent of phonon modes. The origin of this interaction can be understood from spin-dependent displacements of the resonators' equilibrium positions as described in Figure~\ref{fig:Figure2} a).  The evolution under Hamiltonian~\eqref{eq:Hpolaron} then implements spin entangling operations of the form
\begin{equation}\label{eq:Ug}
U_g(t_g)= e^{i (\sum_{i,j} M_{ij}\sigma_z^i \sigma_z^j) t_g}.
\end{equation}
For $N=2$ an initially separable spin superposition state, e.g.,  $|\psi\rangle_0=\prod_{i =1,2} (|0_i\rangle+|1_i\rangle)/\sqrt{2}$ evolves into an entangled state $|\psi\rangle(t_g)= (|00\rangle+|11\rangle +i |01\rangle+i |10\rangle)/2$ on a timescale $t_g=\pi/(4 |M|)$. Here $M=\eta^2 \omega_r(1/\xi-1)/4$ is the characteristic interaction strength, $\eta=\lambda/\omega_r$ is the magnetic coupling parameter and  $\xi=\omega_0/\omega_1$ the ratio between the two phonon frequencies where $\xi(g\ll \omega_r)\simeq 1+2 g/\omega_r$ and $\xi(g\gg \omega_r)\simeq 2  \sqrt{g/\omega_r}$. We see that the gate speed is optimized under strong coupling conditions $\lambda, g\gtrsim \omega_r$, i.e., when the spin displaces the resonator by more than its zero point motion. In principle this condition can always be achieved by choosing a resonator with smaller vibration frequency $\omega_r$, but practical limitations can prevent this as discussed below.

For $N>2$ the coupling matrix $M_{ij}$ depends on the phonon properties and therefore on the layout of the underlying electric circuit (see Figure~\ref{fig:Figure2} b)). For example, connecting neighboring resonators by individually isolated wires results in a nearest neighbor phonon coupling $g_{i,i+1}=g$ and a spectrum $\omega_n=\sqrt{\omega_r^2 \! +\!4 \omega_r g \left(1\!+\!  \cos((n\!+\!1)\pi/N) \right)}$. For $g \ll \omega_r$ we find that  $M_{i,i\pm 1}\simeq \eta^2 g/4$ and interactions quickly decay like $M_{i,i\pm m}\sim (g/\omega_r)^{(m-1)}$ for larger spin separations. Hamiltonian~\eqref{eq:Hpolaron} then corresponds to an Ising model with nearest neighbor coupling $M$ and extensions to a 2D lattice are discussed below. In contrast, by coupling all resonators to a single wire we obtain $g_{ij}= g/(N-1)$  and the phonon spectrum consists of only two frequencies: $\omega_0=\sqrt{\omega_r^2+2 \omega_r g N/(N-1)}$ is the frequency of the center of mass mode with $c_{0,i}=1/\sqrt{N}$, while all other orthogonal modes are unaffected and $\omega_{n>0}=\omega_r$.
This configuration translates into an  `infinite range'  or  collective spin model $H\simeq M S_z^2$, where $S_z=\sum_i \sigma^i_z/\sqrt{N}$, applicable for the generation of spin squeezed~\cite{KitagawaPRA1993} or highly entangled $N$ particle GHz states~\cite{MolmerPRL1999}. Generally we observe that the NEMS array implements a mapping of a given circuit layout onto a corresponding Ising model $H_{\rm Ising}$, and thereby also maps the flexibility of electric circuit design onto an equivalent flexibility in the design of spin-spin interactions.

\section{Decoherence and spin echo}
Our discussion of spin-spin interactions so far has ignored decoherence processes in form of mechanical dissipation and spin dephasing  which degrade the implementation of coherent gate operations in a realistic setting. Before addressing the effect of decoherence on gate operations we remark that dephasing of an idle qubit is eliminated to a large extend by encoding quantum information in nuclear spin degrees of freedom located in the vicinity of the electronic spin.  Prolonged storage times and techniques for the implementation of gate operations between nuclear and electronic spins on a $10-100$ ns timescale have already been demonstrated in several experiments~\cite{MortonNatPhys2006,JelezkoPRL2004,DuttScience2007,MortonNAT2008}. In our setup this approach also allows us to use swap operations between nuclear and electronic spins to switch on and off the Ising interaction for a specific set of qubits and in a controlled way. In a scenario where each electronic spin is coupled to multiple nuclear spins we might in addition benefit from entanglement purification schemes~\cite{DuerPRL2003,LiangPRA2007}, relaxing the bounds on tolerable errors in each individual gate operation.   

\begin{figure}
\begin{centering}
\includegraphics[width=0.65\textwidth]{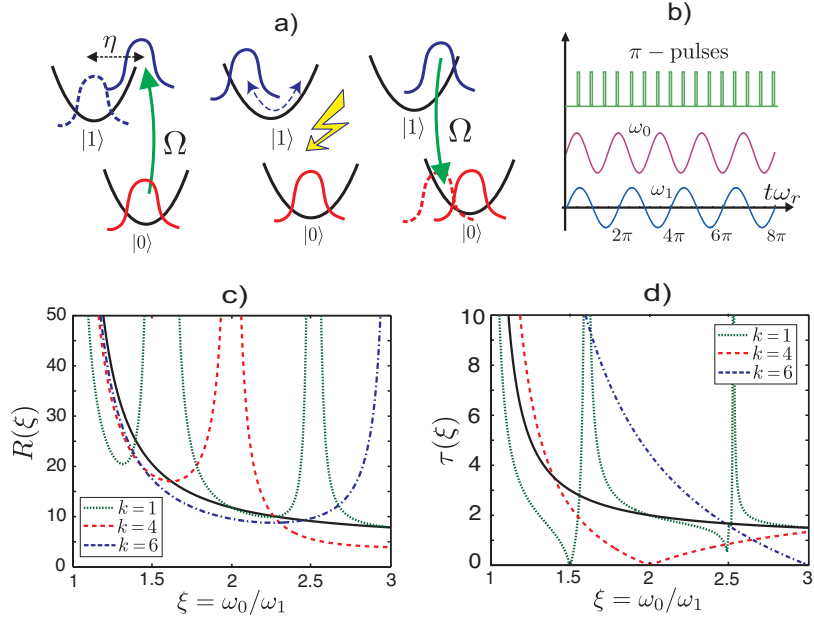}
\caption{{\bf Spin echo and motional dephasing.}
{\bf a}, A fast flip of the spin state is accompanied by a displacement of the resonator state by $\pm \eta$ relative to its equilibrium position 
and entangles spin and motional degrees of freedom. Dephasing of motional superposition states during the gate and a non-vanishing excitations of the phonon modes at the end of the gate sequence reduce the gate fidelity.  {\bf b}, To optimize the gate fidelity pulses are chosen to be commensurate with the oscillation frequency which here is shown for a spin echo sequence with $k=4$ $\pi$-pulses per cycle and a frequency ratio $\xi=\omega_0/\omega_r=5/4$.  {\bf c}, Motional decoherence parameter $R(\xi)$ as used in Eq.~\eqref{eq:ApproxFidelity} without spin echo (solid line) and spin echo sequences with $k=1,4,6$ pulses per oscillation period.  {\bf d}, The normalized gate time $\tau(\xi)$ for the same conditions.} \label{fig:Figure3}
\end{centering}
\end{figure}

The implementation of spin entanglement operations requires the creation of an electronic spin superposition which then evolves under the Ising Hamiltonian~\eqref{eq:Hpolaron} for a time $t_g$.  During this time spins interact with their local magnetic environment which leads to a loss of coherence  $\sim e^{-(t_g/T_2)^\alpha}$, where  $\alpha\geq 1$ depends on specific properties of the environment and the gate sequence~\cite{SousaPRB2003,CoishPRB2004,MazePRB2008}. For solid state spins long coherence times $T_2\sim 1$ ms are typically achieved only in combination with spin echo techniques~\cite{SpinEcho,UhrigPRL2006} where gate operations  are interrupted by a sequence of fast $\pi$ rotations of the spins  to cancel out low frequency noise. In the present setting such techniques serve the additional purpose of reducing the effect of electric $1/f$ noise~\cite{DuttaRMP1981}, which is filtered by the response of the charged resonator and converted into magnetic field fluctuations.  However, as can be seen by the first term in Hamiltonian~\eqref{eq:Hpolaron} any spin rotation, which includes spin echo pulses as well as the initial spin preparation step, is also accompanied by a displacement of the resonator modes and entangles spin and motional degrees of freedom. Specifically, a total gate sequence of $N_p$ echo pulses applied at times $t_{p=1,\dots, N_p}$ will excite each phonon mode by a spin-dependent amplitude proportional to $\beta(\omega_n)=(1-e^{i\omega_nt_g})/2+ \sum_{p=1}^{N_p} (-1)^p e^{i\omega_n t_p}$. As illustrated in Figure~\ref{fig:Figure3} a finite excitation at the end of the gate sequence, $\beta(\omega_n)\neq 0$,  as well as the dephasing of motional superposition states during the evolution then degrade the gate fidelity.

In Methods we detail a model for mechanical decoherence caused by interactions of the resonator modes with a thermal phonon reservoir and derive general expressions for gate fidelities for arbitrary spin echo sequences. For the case of two coupled spins and small gate errors the fidelity of a single entanglement operations can approximately be written as
\begin{equation}\label{eq:ApproxFidelity}
\mathcal{F}\simeq 1 - 4 \eta^2 N_{th} \Delta \beta^2-R(\xi)  \frac{\Gamma_{m}}{\omega_r} - \left(\frac{\omega_r\tau(\xi) }{\lambda^2 T_2}\right)^\alpha.
\end{equation}
Here $N_{th}\simeq k_B T/\hbar \omega_r$  is the  equilibrium occupation number and  $\Gamma_m=k_BT/\hbar Q$ the characteristic motional decoherence rate for a mechanical quality factor $Q$  and a support temperature $T$. The mean excitation amplitude $\Delta \beta =\sqrt{ \frac{1}{N}\sum_{n} (\frac{\omega_r}{\omega_n})^3  |\beta(\omega_n)|^2}$, the dimensionless decoherence parameter $R(\xi)$ and the normalized gate time $\tau(\xi)=t_g/(\omega_r/\eta^2)$ depend on the frequency ratio $\xi=\omega_0/\omega_r$ and on the specific spin echo pulse sequence. For a gate operation without the application of any $\pi$-pulses $R(\xi)= 3 \pi(\xi+\xi^{-1})/ 2(\xi-1)$ and $\tau(\xi)=\pi\xi /(\xi-1)$.

Figure \ref{fig:Figure3} illustrates the basic strategy for obtaining high fidelity gates by avoiding enhanced motional dephasing. Here we have assumed a sequence of $N_p=n\times k$ fast $\pi$ pulses applied at times $\omega_r t_p= p \times 2\pi/k$, where $n$ and $k$ are integers. The condition $\beta(\omega_{0,1})=0$ can then be satisfied by tuning the phonon frequencies such that $\xi=m/n$  is a rational number. Under those assumptions the fidelity is only limited by spin and motional dephasing during the evolution and the resulting values for $R(\xi)$ and $\tau(\xi)$ are plotted in Figure \ref{fig:Figure3} c) and d) for different $k$. Note in particular, that the decoherence parameter $R(\xi)$  exhibits a strong enhancement at frequencies $\omega_0$ near odd multiples of $\omega_r k/2$. Those resonances indicate the excitation of a large resonator superposition during the gate sequence which also strongly decoheres. This can be avoided either for specific values of $\xi$ or in general by choosing a fast pulse sequence with $k > 4 $ where the resonator response is highly suppressed for a large frequency range. By that magnetic and electric low frequency noise can be systematically eliminated without introducing additional mechanical decoherence. While a moderate speed up of gate operations can be obtained for specific parameters we observe for very large values of $k$ an increase of the total gate time $\tau(\xi)\sim k^2$ due to a freezing of the resonator positions by rapid spin flips. However, for a slow magnetic environment it is expected that this effect is compensated by a similar scaling of the spin dephasing time $T_2$~\cite{UhrigPRL2006}. For a specific experimental realization further improvements can be obtained from more advanced spin echo schemes~\cite{UhrigPRL2006} and numerical optimization methods~\cite{GarciaRipollPRA2004}.

\begin{figure}
\begin{centering}
\includegraphics[width=0.75\textwidth]{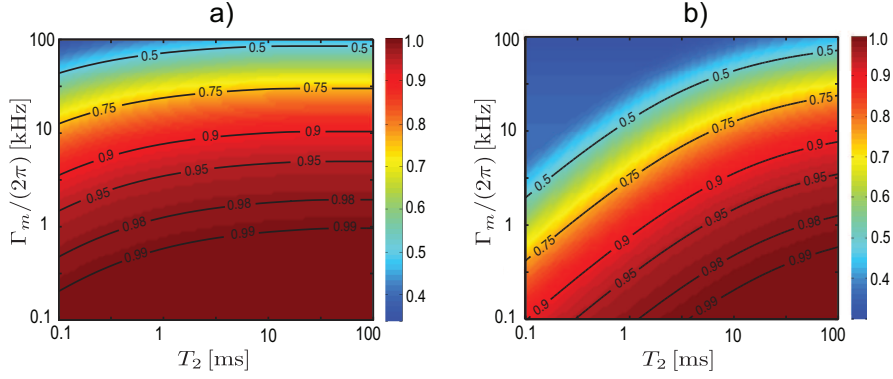}
\caption{{\bf Gate fidelity.} The fidelity $\mathcal{F}$ of a two qubit operation is plotted as a function of the motional decoherence rate $\Gamma_m=k_BT/\hbar Q$ and the spin coherence time $T_2$ assuming a loss of spin coherence $\sim e^{-(t_g/T_2)^3}$.   For a general discussion effects of specific spin echo sequences are neglected and motional dephasing is evaluated for a gate without any additional $\pi$-pulses, which captures well the average dependence. The magnetic coupling strength is {\bf a}, $\lambda/2\pi=100$ kHz and {\bf b}, $\lambda/2\pi=10$ kHz. In both plots $g/2\pi=500$ kHz and the resonator frequency $\omega_r/2\pi$ is optimized within the interval from 1 kHz to 5 MHz. The linear increase of the lines of constant fidelity indicates a scaling of gate errors as $\sim (\Gamma_m/\lambda^2T_2)^{\frac{3}{4}}$ while for increasing $T_2$ the fidelity is limited by motional dephasing, $1-\mathcal{F}\sim  \Gamma_m/g$.} \label{fig:Figure4}
\end{centering}
\end{figure}

The general expression of $\mathcal{F}$ given in Eq.~\eqref{eq:ApproxFidelity} shows that for ideal spin qubits the gate fidelity is independent of $\lambda$. This can be understood from the fact that both the effective spin-spin coupling as well as the dephasing rate of a motional superposition states scale with the square of the displacement amplitude $\eta=\lambda/\omega_r$.
For $\omega_r\gg g$ we find that $R(\xi)\sim \omega_r/g$ and the fidelity is limited by the ratio $\Gamma_m/g$, meaning that strong coupling conditions are required only with respect to electric interactions.  However, the overall gate time $t_g\sim \eta^{-2}$ increases for small $\eta$ and for a finite spin dephasing time $T_2$ there is a competition between spin and motional decoherence processes. Then, for a fixed $\lambda$ and under the assumption that $g \gtrsim \omega_r$  is satisfied there exists an optimal frequency $\omega_r^{op}$ for which gate errors scale as $\sim (\Gamma_m/\lambda^2T_2)^{\frac{\alpha}{\alpha+1}}$ and high gate fidelities can be achieved for either small spin or motional dephasing times. In Figure~\ref{fig:Figure4} we plot numerical values for the optimized gate fidelity as a function of $\Gamma_m$ and $T_2$.  At $T=100$ mK a mechanical quality factor of $Q\sim 10^6$  corresponds to $\Gamma_m/2\pi\simeq 2$ kHz and for $\lambda/(2\pi)=100$ kHz, $g/(2\pi)\simeq 500$ kHz gate fidelities close to  $\mathcal{F}\sim 0.99$ can be achieved with realistic spin coherence times $T_2\approx 1-10$ ms.  Note that in combination with entanglement purification schemes~\cite{DuerPRL2003,LiangPRA2007} fidelities of $\mathcal{F}
> 2/3$ are in principle sufficient, which would relax some of the experimental requirements and makes this technique applicable for temperatures up to $T\sim 1$ K. 

Our discussion so far has ignored the effect of pulse errors which under realistic conditions result in a finite excitation $\Delta\beta\neq 0$ and decrease the fidelity by factor $\sim e^{-4\eta^2 N_{th}\Delta \beta^2}$. For typical equilibrium occupation numbers $N_{th}\sim10^3-10^4$ the gate therefore becomes highly sensitive to any imperfections and limits the application of this technique to small values of $\eta$. However, in Methods we show that the equilibrium occupation number $N_{th}$ in Eq.~\eqref{eq:ApproxFidelity} is replaced by $N_i+1/2$ when the resonator modes are cooled to a lower occupation number $N_i\ll N_{th}$ just before the gate operation. Therefore, already moderate cooling to $N_i\sim 10-100$~\cite{NaikNAT2006,GiganNAT2006,SchliesserNATPHY2008,ThompsonNAT2008,TeufelPRL2008,NVCantilever} makes the gate robust against technical imperfections, while still allowing coupling parameters in the range of $\eta\sim 0.1-1$.

\section{Quantum computing $\&$ scalability}
\begin{figure}
\begin{centering}
\includegraphics[width=0.75\textwidth]{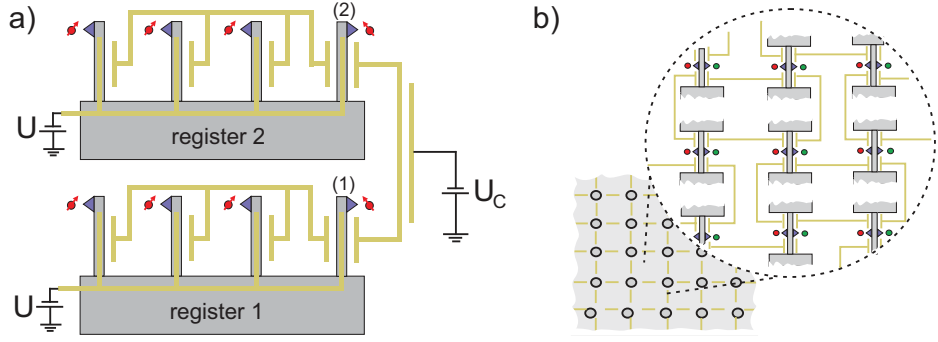}
\caption{{\bf Scalable quantum information processing.} {\bf a}, Implementation of a switchable coupling between two sub-registers. 
For a control voltage $U_c=0$ a finite electric coupling $g_{12}\approx g$ between resonator 1 and 2 enables gate operations between spins located in different registers. For $U_c\simeq U$ there is no energy associated with charge flowing from the gate capacitors onto the wire and thereby the coupling $g_{12}\simeq 0$  is switched off. {\bf b}, A scalable quantum computing architecture, based on the concept of one-way quantum computing (OWQC)~\cite{RaussendorfPRL2001}. Resonators are ordered on a 2D lattice and coupled to its four neighbors electrostatically. This configuration results in an effective spin Hamiltonian $H\simeq M \sum_{\langle i,j\rangle} \sigma_z^i\sigma_z^j$, which can be used to generate cluster states as an entanglement resource.  With each node consisting of multiple spins, the cluster state can be stored in long lived electronic or nuclear spin qubits (green dots) while local measurements, the actual computation step in OWCQ, can be performed via an optical active spin (red dots), e.g.,  an NV center. } \label{fig:Figure5}
\end{centering}
\end{figure}

We finally discuss potential realizations of scalable quantum computing architectures which are based on NEMS mediated spin-spin interactions. Let us first consider a small sub-unit of $N_s$ resonators coupled by a single wire. In that case the phonon spectrum is independent of $N_s$ and the pulse sequences discussed above directly apply for any two and multi qubit gate. For state preparation and detection we distinguish between at least one ``control" and the remaining ``passive" qubits. For the control qubit we choose, for example, an NV center, which can be polarized and detected optically. Two qubit gates are used to perform a mapping $|0\rangle_c(\alpha|0\rangle_i+\beta|1\rangle_i)\rightarrow   (\alpha |0\rangle_c|0\rangle_i+\beta |1\rangle_c|1\rangle_i)$ such that a successive detection of the control spin implements a QND measurement of the state of the $i$-th spin. For state preparation the known spin state is rotated afterwards to the desired target state using local operations.  By that approach the ``passive" spins can be optimized with respect to their coherence properties, optical detection is spatially separated from storage qubits and optical pumping of the control spin can be employed to cool the phonon modes between gate operations~\cite{NVCantilever}. The scaling of Ising interactions $\sim 1/N_{s}$ however limits the size of a single sub-register to a few or, including several electronic/nuclear spins per resonator,  to a few tens of qubits. To go beyond this limit individual sub-registers can be connected by a switchable coupling as described in Figure~\ref{fig:Figure5} a).
Although the mode spectrum of two coupled registers is slightly more complicated (e.g. 5 different frequencies) the operation of a large scale quantum computer can still be reduced to gate operations within one or at most two sub-registers.

An alternative approach is sketched in Figure~\ref{fig:Figure5} b) where the resonators are ordered in a 2D lattice structure with nearest neighbor coupling. This layout is motivated by the concept of one way quantum computing~\cite{RaussendorfPRL2001} where in a first step a cluster state is prepared as an entanglement resource while the actual computation is done by measurements and local spin operations only. The cluster state is generated by applying the gate operation $U_g$ given in Eq.~\eqref{eq:Ug} consecutively or in parallel to each spin $i$ and its 4 nearest neighbors. The nearest neighbor coupling $M_{i,i+1}\simeq \eta^2 g/4$ is independent of the lattice size and since interactions  decay quickly with distance many gates can be carried out in parallel. However, for a large lattice the phonon spectrum is almost continuous and the condition $\beta(\omega_n)=0$ can no longer be strictly fulfilled and $\Delta \beta >0$.
A numerical evaluation of $\Delta \beta$ for $g/\omega_r=0.2$ and a $k=4$ spin echo sequence commensurate with ${\rm max}\{\omega_n\}$ gives $\Delta \beta^2\sim0.015$, roughly independent of the length of the gate sequence. For $\eta^2N_i < 1$ the resulting error of $\sim1$ \% is still sufficiently low.

\section{Conclusions $\&$ Outlook}
In summary we have proposed the application of NEMS as a universal quantum transducer for spin-spin interactions. Compared to direct magnetic coupling or probabilistic optical entanglement schemes our approach enables the implementation of long-range and deterministic spin entanglement operations as well as the design and control of multi-spin interactions by simple electric circuitry.  The universality of the basic underlying concept, namely to use the mechanical resonator for a coherent conversion of magnetic into electric dipoles, also opens a wide range of possibilities for the integration of electronic and nuclear spins with other charge based quantum systems. Specifically, it might be interesting to consider hybrid architectures by coupling spins with transmission line cavities~\cite{TeufelPRL2008}, charge qubits~\cite{ArmourPRL2002} and trapped ions~\cite{TianPRL,HensingerPRA2005} or atoms~\cite{TreutleinPRL2007}. Moreover, this techniques can be applicable for a remote magnetic sensing of ``dark" spins in a condensed matter or biological environment which is incompatible with direct laser illumination.  In a broader perspective the quantum transducer ability of NEMS can therefore be seen as one of the fundamental applications of ``quantum'' mechanical systems.

\section{METHODS}
To evaluate the effect of mechanical dissipation we consider a model described by a total Hamiltonian 
\begin{equation}\label{eq:HEnvironment}
H_{\rm tot}= H+ \sum_{n,k} \frac{g_{k}}{2} \left(a_n+a_n^\dag\right)\left(b_{n,k}+b^\dag_{n,k}\right) + \sum_{n,k,} \omega_k b_{n,k}^\dag b_{n,k},
\end{equation}
where $H$ is the system Hamiltonian~\eqref{eq:Model} and $b_{n,k}$ are bosonic bath operators, $[b_{n,k},b^\dag_{n',k'}]=\delta_{nn'}\delta_{kk'}$. The environment is characterized by the spectral density $J(\omega)=\frac{\pi}{2}\sum_{k} g_k^2\delta(\omega-\omega_k)$~\cite{SpinBosonRev}, which we assume to be equal for all modes. Clamping losses can be described by a purely ohmic environment $J(\omega)=\omega/Q$, while the effect of electric 1/f noise can be taken into account by setting $J(\omega)=const.$ in Eq.~\eqref{eq:Fn} given below. In the limit of fast $\pi$ pulses the bare evolution of the spin operators reduces to $\sigma^i_z(t)=f(t)\sigma^i_z$. Here $f(t)= 2\sum_{p=0}^{N_p+1} z_p \theta(t-t_p)$, $\theta(t)$ is the unit step function, $z_0=-z_{N_p+1}=1/2$ and $z_p=(-1)^p$ otherwise. It describes the effect of alternating spin flips at times $t_p$ and the first and last term account for the inital and final spin preparation step at $t_0=0$ and $t_{N_p+1}=t_g$. The total Hamiltonian can then be written as
\begin{equation}\label{eq:Hint}
H_{\rm tot}=  H_{\rm osc} + \frac{\lambda}{2} \sum_{n} f(t) x_n  S_z^n, 
\end{equation}
where $S_z^n=\sum_i c_{n,i} \sigma_z^i$, $x_n=(a_n+a^\dag_n)$ and $H_{\rm osc}$ is the Hamiltonian of the coupled resonator and bath degrees of freedom.  By changing to the interaction picture, $x_n(t)=e^{i H_{\rm osc} t}x_n e^{-i H_{\rm osc} t}$, the evolution generated by Hamiltonian~\eqref{eq:Hint} is
\begin{equation}
U_g(t_g)=  \prod_{n=1}^{N} e^{-i\frac{\lambda}{2}\int_0^{t_g}ds f(s)x_n(s)S_z^n} e^{i \Phi_n (S_z^n)^2},
\end{equation}
where the geometric phases $\Phi_n$ are given by
\begin{equation}\label{eq:Phases}
\Phi_n= i \frac{\lambda^2}{8} \int_0^{t_g} ds \int_0^s ds'  f(s)f(s')[x_n(s),x_n(s')]\,.
\end{equation}
Eq.~\eqref{eq:Phases} is evaluated by re-expressing $x_n(t)$ in terms of eigen-operators $d_{n,k}$ and eigenfrequencies $\omega_{n,k}$ of the coupled oscillator Hamiltonian $H_{\rm osc}$, i.e. $\lambda x_n(t)= \sum_k \lambda_{n,k} (d^\dag_{n,k}e^{i\omega_{n,k}t}+d_{n,k}e^{-i\omega_{n,k}t})$. After an integration by parts we write the result as
\begin{equation}\label{eq:PhaseExact}
\Phi_n(t_g)= \frac{\eta_n^2}{2\pi }  \int_0^\infty d\omega \,J_{\rm eff}^n(\omega) \left[ \frac{t_g}{\omega} + \frac{4}{\omega^2} \sum_{p=2}^{N_p}\sum_{p'=1}^{p-1}  z_pz_{p'} \sin(\omega(t_p-t_{p'}))\right].
\end{equation}
Here $\eta_n=\lambda/\omega_n$ and $ J_{\rm eff}^n(\omega):=\frac{\pi}{2 \eta_n^2}\sum_k \lambda_{n,k}^2\delta(\omega-\omega_{n,k})$ can be expressed in terms of $J(\omega)$ by the relation $J_{\rm eff}^n (\omega)=  J(\omega) \omega_n^4 /[ (\omega_n^2-\omega^2)^2+\omega_n^2 J^2(\omega)]$~\cite{Garg1985}. Evaluating Eq.~\eqref{eq:PhaseExact} for $Q\gg1$ we recover from the first term in brackets the bare Ising interactions, $\Phi_n(t_g)= \eta^2 \omega_n t_g/4$, plus small corrections $\mathcal{O} (t_g\omega_r/Q)$. The second term in Eq.~\eqref{eq:PhaseExact} describes additional geometric phases which depend on the pulse sequence and modify the effective spin coupling strength, which, e. g., for two spins we define as $M_{\rm eff}=(\Phi_0-\Phi_1)/t_g$. For spin echo sequences discussed in the main part of the paper we find a significant enhancement of  $M_{\rm eff }$ only in combination with strong motional decoherence, while $M_{\rm eff}\sim 1/k^2$ in the limit of fast echo pulses $k\gg 1$.

For a given initial pure spin state $|\psi_0\rangle$ and a target state $|\psi_f\rangle=\prod_n e^{i\Phi_n(S_z^n)^2}|\psi_0\rangle$, we define the gate fidelity as $
\mathcal{F}= {\rm Tr}\{ \langle \psi_f | U_g \big( |\psi_0 \rangle\langle\psi_0|\otimes \rho_{\rm osc}(0)\big)U_g^\dag|\psi_f\rangle \}
$, where $\rho_{\rm osc}(0)$ is the initial state of the oscillator modes. We here assume pure dephasing processes only and write $|\psi_0\rangle=\sum_{\vec s} c_{\vec s}|{\vec s}\rangle$ where $\sigma_z^i|s_i\rangle = s_i|s_i\rangle$. Then, by setting $s_n=\sum_i c_{n,i} s_i$ the fidelity can be written as 
\begin{equation}\label{eq:Fidelity}
\mathcal{F} = \sum_{\vec s, \vec r} | c_{\vec s}|^2 |c_{\vec r}|^2 e^{-\frac{1}{4}\sum_n F_n (s_n- r_n)^2}.
\end{equation}
The effect of motional dephasing of each collective mode $n$ is expressed in terms of coefficients $F_n$, which we evaluate in the following. For a symmetric two qubit state with $c_{\vec s}=1/4$, in the limit of high fidelities and a total gate time $t_g\simeq \pi/(4M_{\rm eff})$  we obtain $\mathcal{F}\simeq 1- \pi \Gamma_{\rm eff}/(4 M_{\rm eff})-F_s(t_g)$. Here we have added the bare spin dephasing $F_s(t_g)\simeq (t_g/T_2)^\alpha$ and introduced an average motional dephasing rate $\Gamma_{\rm eff}=(F_0+F_1)/(2t_g)$.
With the definition $R(\xi):= \pi \omega_r \Gamma_{\rm eff}/(4 M_{\rm eff} \Gamma_m)$ we obtain the result given in Eq.~\eqref{eq:ApproxFidelity}. 

  When the resonator modes are in thermal equilibrium with the environment $\rho_{\rm osc}(0)= \prod_{n,k}\rho_{n,k}$ is simply a product of thermal states for each mode $d_{n,k}$. Therefore, we can also decompose $x_n(t)$ into eigen-operators and evaluate the thermal expectation values of $U_g$ for each mode. We obtain  
\begin{equation}\label{eq:Fn}
F_n=  \frac{4 \eta_n^2}{\pi}\int_0^\infty d\omega \frac{J_{\rm eff}^n(\omega)}{\omega^2} \coth\left(\frac{\hbar \omega}{2k_BT}\right)\left|\beta(\omega) \right|^2\,. 
\end{equation}
Here $|\beta(\omega)|^2=\sin^2(\omega t_g/2)$ for a gate without spin echo and  $|\beta(\omega)|^2=\sin^2(n\pi\omega/\omega_r)\tan^2(\pi\omega/(k\omega_r))$ for $k$ equidistant spin echo pulse per oscillation period $2\pi/\omega_r$.
Eq.~\eqref{eq:Fn} is familiar from discussion of spin dephasing within the spin boson model (see e.g.~\cite{SpinBosonRev, UhrigPRL2006}), but here qualitatively different results emerge from the resonant structure of $J_{\rm eff}^n(\omega)$.
For isolated resonator modes, $J(\omega)\rightarrow 0$,  Eq.~\eqref{eq:Fn} reduces to $F_n\simeq 4 \eta_n^2 N^{(n)}_{th} |\beta(\omega_n)|^2$ as a result of the residual entanglement between spins and the bare resonator modes at the end of the gate sequence.  For the low frequency part of the integral in Eq.~\eqref{eq:Fn} we can approximate $J_{\rm eff}^n(\omega\ll \omega_n )\approx J(\omega)$ and for an ohmic bath and without spin echo we obtain $F_n^l \simeq 2\eta_n^2 \Gamma_m t_g$. Including spin echo we find that $\lim_{\omega\rightarrow 0} J(\omega)|\beta(\omega)|^2/\omega^3=0$ both for ohmic and $1/f$ noise and in the latter case $F_n^l$ grows only logarithmically with $t_g$.  The remaining contribution to $F_n$ then comes mainly form near resonant modes $\omega\approx \omega_n$. 

If the phonon modes are pre-cooled to a temperature $T_i \ll T$ the initial density operator $\rho_{\rm osc}(0)=\prod_{n} \rho_n(T_i)\prod_k \rho_{n,k}(T)$ is diagonal in the uncoupled resonator and bath operators. In that case we calculate the time evolution for the phonon mode operators $x_n(t)$ which for $t\omega_r \gg1$ obey the equation of motion,
\begin{equation}\label{eq:Xn}
\ddot x_n(t)+\gamma_n \dot x_n(t) +\omega_n^2 x_n(t)=- \omega_n\sum_k g_k\left(b^\dag_{n,k} e^{i\omega_kt} +b_{n,k} e^{-i\omega_k t}\right),
\end{equation}
where $\gamma_n=\omega_n/Q$. The solutions of Eq.~\eqref{eq:Xn} can be divided in terms containing system and bath operators only, $x_n(t)=x_n^{(s)}(t)+x_n^{(b)}(t)$,  and accordingly we decompose $F_n=F_n^{(s)}+F_n^{(b)}$.  For a weakly damped resonator we obtain $x_n^{(s)}(t)\simeq(x_n\cos(\omega_n t)+p_n\sin(\omega_n t))e^{-\gamma_n t/2}$ and  $F_n^{(s)}\simeq  2 \eta_n^2 (2N_i+1)|\beta(\omega_n+i\gamma_n/2)|^2$, where $N_i$ is the initial occupation number. The contribution from the bath is
\begin{equation}
x_n^{(b)}(t)=- \omega_n\sum_k g_k \left( v_n(\omega_k,t) b^\dag _{n,k} +v^*(\omega_k,t)b^\dag _{n,k} \right),
\end{equation}
where $v_n(\omega,t)=\mathcal{L}^{-1}[(s^2+\gamma s+\omega_n^2)^{-1}(s-i\omega)^{-1}]$ and  $\mathcal{L}^{-1}$ denotes the inverse Laplace transformation.  For low frequencies, $\omega_k<\omega_r$ we approximate $v_n(\omega,t)\approx e^{i\omega t}/\omega_n^2$, and in this regime we recover the same result as given in Eq.~\eqref{eq:Fn} for the equilibrium case. For near-resonant modes 
we use $
v_n(\omega,t)\approx (e^{i\omega t}-e^{(i\omega_n-\gamma_n/2) t})/(\omega_n(2(\omega_n-\omega)+i\gamma_n))$ and obtain
\begin{equation}
F_n^{(b)}\simeq F^l_n+   \eta_n^2 N_{th}^{(n)}\frac{2}{\pi}\int_{-\infty}^{\infty} d\omega \frac{ \gamma_n |\beta(\omega_n+i\gamma_n/2)-\beta(\omega)|^2}{(\omega-\omega_n)^2+\gamma_n^2/4}\,.
\end{equation}
Evaluating this integral we find in summary that for $N_i\ll N_{th}^{(n)}$ and  $\gamma_n t_g \ll 1$ the motional decoherence coefficients are  
\begin{equation}
F_n\simeq  F^l_n+ 4\eta_n^2 \left[ \left(N_i+\frac{1}{2}\right)|\beta(\omega_n)|^2 + \Gamma_m \sum_{p,p'} z_p z_{p'} e^{i\omega_n(t_p-t_{p'})} |t_p-t_p'|\right] \,.
\end{equation}
We see that to first order in $\gamma_n t_g$ the effect of pulse errors, $\beta(\omega_n)\neq0$, scales only with the initial occupation number  $N_i$ while low frequency noise and interactions with near resonate environmental modes leads to dephasing proportional to the bath temperature, $\Gamma_m=\gamma_nN_{th}^{(n)}$. Similar conclusion can be derived from a master equation approach~\cite{GarciaRipollPRA2004}, which however does not treat low frequency noise correctly and ignores the high frequency `cutoff' $v_n(\omega\gg \omega_n,t)\sim \omega^{-2}$.

\section{Acknowledgements}

We gratefully acknowledge discussion with M. Aspelmeyer and K. Schwab.  This work is supported by ITAMP, NSF, CUA,
DARPA, and the Packard Foundation. P. Z. acknowledges support by SFB FOQUS and EU Networks.
Correspondence and requests for materials should be addressed to P. R.






\end{document}